\documentclass[12pt,fleqn]{article}

\usepackage[letterpaper,margin=1in]{geometry}
\usepackage{amsmath,amssymb,graphicx,caption,float}
\usepackage[affil-it]{authblk}

\numberwithin{equation}{section}

\newcommand{\ex}[1]{\textrm{e}^{#1}}
\newcommand{\diff}{\mathrm{d}}
\newcommand{\xcl}{x_{cl}}
\newcommand{\dotxcl}{\dot x_{cl}}
\newcommand{\Lo}{L^{(0)}}

\begin{document}

\title{Generalizations to Feynman's Path Integration Methods in One Dimension}
\author{John W. Russell}
\affil{Department of Physics and Astronomy\\
	University of Hawaii at Manoa\\
	Honolulu, HI 96822, USA}
\date{}
\maketitle

\begin{abstract}
	This paper reviews and generalizes Feynman's path integration methods which use time slicing with straight line segments and Fourier sine series. The generalizations are done from variational calculus considerations and in one dimension for simplicity in demonstrating concepts.
\end{abstract}

\section{Introduction}

	When a time-independent Hamiltonian has an energy eigenvalue spectrum $\{E_n\}_{n\in\mathbb N}$, a \textit{propagator} $K(\pmb x_b, t_b; \pmb x_a, t_a)$ can be used to ``propagate" a wavefunction $\psi(\pmb x,t)$ comprised of the corresponding eigenfunctions $\{\varphi_n(\pmb x)\}_{n \in \mathbb N}$,
    \[\psi(\pmb x, t)=\sum_{n \in \mathbb N} c_n \varphi_n(\pmb x) \ex{-i E_n t / \hbar}, \qquad \sum_{n \in \mathbb N} \lvert c_n\rvert^2=1,\]
    from a point $a$ in space and time to another point $b$,
    \begin{equation}
    	\psi(\pmb x_b, t_b)=\int_{\mathbb R^3} \! \mathrm{d}^3 x_a \, K(\pmb x_b, t_b; \pmb x_a, t_a) \psi(\pmb x_a, t_a).
    	\label{wavefnexpand}
    \end{equation}
    The propagator is expandable in this basis of eigenfunctions as
	\begin{equation}
        K(\pmb x_b, t_b; \pmb x_a, t_a)=\sum_{n \in \mathbb N} \varphi_n(\pmb x_b) \varphi_n^*(\pmb x_a) \ex{-i E_n (t_b - t_a) / \hbar}.
        \label{inbasisprop}
	\end{equation}
    Therefore, if the propagator can be determined before explicitly finding eigenfunctions, \eqref{inbasisprop} can be used to determine the energy eigenvalue spectrum. This can be achieved with path integral formalism.
    
    The outline of the paper is as follows. Material is reviewed in the following order: variational calculus, the path integral representation of the propagator, Feynman's time slicing path integration method, and then his Fourier series method \cite{FeynmanHibbs}. With the reviews serving to introduce notation and to provide justification, generalizations to these two methods are then presented. We shall be considering propagators for one dimensional systems.
    
\section{Variational calculus}

    Variational calculus uses objects known as \textit{functionals}, functions that map input vector functions onto output scalars.  A goal in the development of variational calculus was to answer the question, ``What functions extremize the value of a functional?" Like a function $f = f(t)$ having extremal value when evaluated at a point $t = t_0$ where its derivative is zero ($f'(t_0) = 0$), a functional $F = F[f]$ is said to have extremal value when evaluated with a function $f = f_0$ resulting in a variation of zero ($\delta F[f_0; g] = 0$). Expressing arbitrary $f$ as $f = f_0 + \sigma g$, where $\sigma$ is a time independent parameter and $g = g(t)$ is an arbitrary continuously differentiable function, variation can be represented with the \textit{G\^{a}teaux differential} \cite{Sagan},
    \begin{equation}
        \delta F[f_0; g] \equiv \left.\frac{\diff}{\diff \sigma}F[f_0 + \sigma g]\right\rvert _{\sigma = 0}.
		\label{1stvar}
	\end{equation}
    To further interpret the meaning of this quantity we provide ourselves the following example of the action from classical mechanics.
    
    In physics, functional extremization is done for functionals relating to quantities such as energy, entropy, and path traveled by light. The energy functional is the \textit{action}, commonly denoted in one dimension as
    \begin{equation}
		S[x] = \int_{t_a}^{t_b} \! \diff t \, L(\dot x, x, t),
		\label{action}
	\end{equation}
	where the integrand $L = L(\dot x, x, t)$ is commonly referred to as the \textit{Lagrangian}. Evaluating the variation of the action for $x = \xcl + \sigma \eta$, we have by chain rule
    \begin{align}
        \delta S[\xcl; \eta] =& \left.\frac{\diff}{\diff \sigma} S[\xcl + \sigma \eta]\right\rvert_{\sigma = 0} \\
        =& \int_{t_a}^{t_b} \! \diff t \left.\frac{\diff}{\diff \sigma} L(\dotxcl + \sigma \dot \eta, \xcl + \sigma \eta, t)\right\rvert_{\sigma = 0} \\
        =& \int_{t_a}^{t_b} \! \diff t \left(\dot \eta \frac{\partial}{\partial \dotxcl} + \eta \frac{\partial}{\partial \xcl}\right) L (\dotxcl, \xcl, t).
        \label{preE-L1}
	\end{align}
    With this last statement, we now discuss the significance of functions $\eta = \eta(t)$ and $\xcl = \xcl(t)$. Such functions $\eta$ exist in what is known as the \textit{space of variations}, satisfying the condition $\eta(t_a) = \eta(t_b) = 0$ \cite{Sagan, Shankar}. With such $\eta$, integration by parts of equation \eqref{preE-L1} leads to
    \begin{equation}
        \delta S[\xcl; \eta] = \int_{t_a}^{t_b} \! \diff t \, \eta \left(-\frac{\diff}{\diff t} \frac{\partial L}{\partial \dotxcl} + \frac{\partial L}{\partial \xcl}\right).
		\label{preE_L2}
	\end{equation}
    If the variation is to be set equal to zero, we then have the \textit{Euler-Lagrange equation},
    \begin{equation}
        -\frac{\diff}{\diff t} \frac{\partial L}{\partial \dotxcl} + \frac{\partial L}{\partial \xcl} = 0.
		\label{E-L}
	\end{equation}
    Functions $\xcl$ which satisfy \eqref{E-L} then extremize the functional. These functions may correspond to minima, maxima, or critical point values of the action. In this context to the action, $\xcl$ is often referred to as the \textit{classical path}.
    
    Like the second derivative test, where a point $t=t_0$ satisfying $f'(t_0) = 0, \ f''(t_0) > 0$ corresponds to a minimum, we find a functional $F[f]$ evaluated with function $f = f_0$ is minimized when $\delta F[f_0; g] = 0,\ \delta^2 F[f_0; g] > 0$, where
	\begin{equation}
        \delta^2 F[f_0; g] \equiv \left.\frac{\diff^2}{\diff \sigma^2} F[f_0 + \sigma g]\right\rvert_{\sigma = 0}
        \label{2ndvar}
	\end{equation}
	is referred to as the \textit{second variation}, and such $f_0$ are called \textit{stable}. In the instance of the action, this becomes
	\begin{align}
		\delta^2 S[\xcl; \eta] =& \left.\frac{\diff^2}{\diff \sigma^2} S[\xcl + \sigma \eta]\right\rvert_{\sigma = 0} \\
        =& \int_{t_a}^{t_b} \! \diff t \left(\dot \eta \frac{\partial}{\partial \dotxcl} + \eta \frac{\partial}{\partial \xcl}\right)^2 L(\dotxcl, \xcl, t)
        \label{preS-L1} \\
        =& \int_{t_a}^{t_b} \! \diff t \left(\dot \eta^2 \frac{\partial^2 L}{\partial \dotxcl^2} + 2 \dot \eta \eta \frac{\partial^2 L}{\partial \dotxcl \partial \xcl} + \eta^2 \frac{\partial^2 L}{\partial \xcl^2}\right) \\
        =& \int_{t_a}^{t_b} \! \diff t \, \eta \left[-\frac{\diff}{\diff t}\left(\dot \eta \frac{\partial^2 L}{\partial \dotxcl^2}\right) + \eta \left(-\frac{\diff}{\diff t} \frac{\partial^2 L}{\partial \dotxcl \partial \xcl} + \frac{\partial^2 L}{\partial\xcl^2}\right)\right].
        \label{preS-L2}
	\end{align}
    This gives rise to the \textit{Legendre condition} \cite{Sagan, GelfandFomin, Schulman},
	\begin{equation}
        \frac{\partial^2 L}{\partial \dotxcl^2} \ge 0, \qquad t \in [t_a, t_b],
		\label{Lgcond}
	\end{equation}
    a necessary condition for the action $S$ to be a minimum over a given $\xcl$ found with \eqref{E-L} when evaluated along $t\in[t_a, t_b]$.
	
\section{The path integral}

    In path integral formalism, the one dimensional propagator can be stated as
	\begin{equation}
        K(x_b, t_b; x_a, t_a)=\int_{x(t_a) = x_a}^{x(t_b) = x_b} \! \mathcal D x \, \ex{i S[x] / \hbar},
		\label{pathintprop}
	\end{equation}
    where $S[x] = S[b,a]$ is the action between endpoints $a$ to $b$ along a given path $x$, and $\mathcal D x$ is a measure of possible paths between those endpoints. The action along $\xcl$ can be denoted as $S[\xcl]$ or $S_{cl}[b,a]$. Common characterizations of these other possible paths and corresponding representations of $\mathcal Dx$ are presented in the following sections.
	
    The relation of the action $S$ to \eqref{pathintprop} is that of a phase. Each path contributes to this phase, and can constructively or destructively interfere with the contribution from the other paths. Paths closer to (further from) $\xcl$ constructively (destructively) interfere \cite{Shankar}. The overall result is that $\xcl$ should represent the most coherent, probable path in propagating a system from points $a$ to $b$. Relating this to the minimization property of stable $\xcl$, we could say smaller action implies longer coherent times. Perhaps some sort of energy-time uncertainty relation can be formulated from this, with interpretation similar to that proposed by Eberly and Singh \cite{EberlySingh}.
	
\section{Feynman's framework of the time slicing method}

    In his book coauthored with Hibbs, Feynman first introduces the time slicing method to the path integral \cite{FeynmanHibbs}. For simplicity, they begin by considering Lagrangians of the form
	\begin{equation}
		L = \frac{1}{2} m \dot x^2 - V(x).
	\end{equation}
    By \textit{time slicing}, what is meant is the action is uniformly partitioned over $t \in [t_a , t_b]$ in $N \in \mathbb N$ time steps $t_n$,
	\begin{equation}
		x_n = x(t_n) , \qquad t_n = t_a + n \epsilon, \qquad \epsilon = (t_b - t_a) / N, \qquad n = 0, 1, \dotsc, N.
		\label{slicingconds}
	\end{equation}
    We are at point $a$ when $n=0$ and at point $b$ when $n = N$. This results in the action being expressible as
	\begin{equation}
        S[b, a] = \sum_{n = 1}^N S[n, n - 1], \qquad S[n, n - 1] = \int_{t_{n - 1}}^{t_n} \! \diff t \, L.
		\label{actpart}
	\end{equation}
    Feynman and Hibbs state the actions $S[n, n - 1]$ under this partition are
	\begin{equation}
        S[n , n - 1] \equiv \frac{1}{2 \epsilon} m(x_n - x_{n - 1})^2 - \epsilon \bar V[n, n - 1],
		\label{shortactFH}
	\end{equation}
    where $\bar V=\bar V[n, n - 1]$ is some average of the potential over $t \in[t_{n - 1}, t_n]$. Throughout their work, Feynman and Hibbs use $\bar V = V\left((x_n + x_{n-1}) / 2\right), \ V(x_{n - 1}),$ or $(V(x_n) + V(x_{n - 1})) / 2$ \cite{FeynmanHibbs}.
	
	Just as a wavefunction can be convolved with a propagator to move it forward in position and time as in \eqref{wavefnexpand}, so too can the propagator with itself,
	\begin{equation}
		K(x_b, t_b; x_a, t_a) = \int_{\mathbb R} \! \diff x_c \, K(x_b, t_b; x_c, t_c) K(x_c, t_c; x_a, t_a).
	\end{equation}
	This is repeatable any number of times, even up to the time sliced intervals of \eqref{slicingconds},
    \begin{equation}
       	 K(x_b, t_b ; x_a, t_a) = \int_{\mathbb R^{N - 1}} \! \left(\prod_{n = 1}^{N - 1} \diff x_n\right) K(x_b, t_b; x_{N - 1}, t_{N - 1}) \dotsm K(x_1, t_1; x_a, t_a).
    	\label{shortpropexpand}
    \end{equation}
    As $N \to \infty$, each convolved propagator approaches the \textit{short-time propagator} \cite{Makri:ST, Makri:FP},
    \begin{equation}
    	K(x_n, t_n; x_{n - 1}, t_{n - 1}) \to \frac{1}{A} \ex{i S[n, n - 1] / \hbar},
    	\label{shortprop}
    \end{equation}
    where $S[n, n - 1]$ is as in \eqref{shortactFH} and $A$ is a normalization constant pertaining to the classical action of the free particle ($S_{cl}[n, n - 1] = m(x_n - x_{n - 1})^2 / 2 \epsilon$),
    \begin{equation}
    	A = \int_\mathbb R \! \diff x_{n - 1} \, \ex{i S_{cl}[n, n - 1] / \hbar} = \left(\frac{i 2 \pi \hbar \epsilon}{m}\right)^{1 / 2}.
		\label{freepartnorm}
	\end{equation}
	
\begin{figure}[H]
		\centering
		\includegraphics[scale = 0.8]{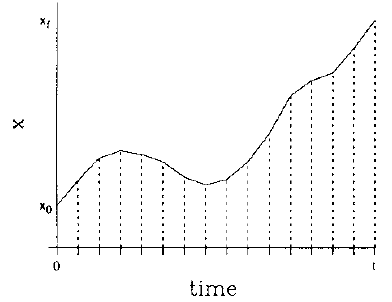}
		\caption{Trajectory time slicing. Over a time $t \in [t_0, t_f]$, the path $x = x(t)$ from point $x_0$ to point $x_f$ in a free particle's action is approximated by straight line segments \cite{Makri:FP}.}
        \label{timeslice}
	\end{figure}

	We can now interpret the measure $\mathcal D x$ in \eqref{pathintprop}. By partitioning the action like Feynman and Hibbs in equation \eqref{shortactFH}, we have a way to express multiple paths from points $a$ to $b$ using straight line segments like in figure \ref{timeslice}. Taking the limit $N \to \infty$ and substituting \eqref{shortprop} into \eqref{shortpropexpand}, with the notation of \eqref{actpart} we arrive at the expression
	\begin{equation}
		K(x_b, t_b; x_a, t_a) = \lim_{N \to \infty} \frac{1}{A} \int_{\mathbb R^{N - 1}} \! \left(\prod_{n = 1}^{N - 1} \frac{\diff x_n}{A}\right) \ex{i S[b, a] / \hbar}.
	\end{equation}
	
\section{Feynman's framework of the ``series" method}

	For actions with Lagrangians comprised of terms $(\dot x, x)$ up to quadratic order \cite{FeynmanHibbs, Shankar}, we may express the corresponding propagator as
	\begin{equation}
		K(x_b, t_b; x_a, t_a) = F(t_b - t_a) \ex{i S_{cl}[b, a] / \hbar}.
		\label{quadprop}
	\end{equation}
	Although \eqref{quadprop} is Gaussian and therefore $F = F(t_b - t_a)$ can be determined much in the same way as $A$ in \eqref{freepartnorm}, Feynman and Hibbs present an alternative method to time slicing to determine $F$ \cite{FeynmanHibbs}. Likely from considering the space of variations, they consider paths which are the sum of the classical path $\xcl$ and paths $y$ which vanish at points $a$ and $b$ ($x = \xcl + y$). With such quadratic Lagrangians, the corresponding action $S[b, a] = S[x] = S[\xcl + y]$ can be expanded up to second variation \eqref{2ndvar}. We then find
	\begin{align}
		S[\xcl + y] =& \int_{t_a}^{t_b} \! \diff t \, L(\dotxcl + \dot y, \xcl + y, t)
		\label{actexpand} \\
		=& \int_{t_a}^{t_b} \! \diff t \, L(\dotxcl, \xcl, t) + 0 + \int_{t_a}^{t_b} \! \diff t \, L(\dot y, y, t)
		\nonumber \\
		=& \; S[\xcl] + S[y].
		\nonumber
	\end{align}
	Transforming $\mathcal D x \mapsto \mathcal D y$ using substitution under integration, $F$ is then expressible as
	\begin{equation}
		F(t_b - t_a) = \int_{y(t_a) = 0}^{y(t_b) = 0} \! \mathcal D y \, \ex{i S[y] / \hbar}.
		\label{Fnorm}
	\end{equation}
	Rather than describing paths $y$ using straight line segments as done in their time slicing method, Feynman and Hibbs represent $y$ using Fourier sine series,
	\begin{equation}
		y(t) = \sum_{n = 1}^\infty a_n \sin\left(n \pi \frac{t - t_a}{t_b - t_a}\right) = \sum_{n = 1}^\infty a_n (-1)^{n - 1} \sin\left(n \pi \frac{t_b - t}{t_b - t_a}\right).
		\label{Fourierseries}
	\end{equation}
	As $N \to \infty$, path integration can be achieved by integrating over the amplitudes $\{a_n\}_{n = 1}^{N - 1}$ rather than the positions $\{y_n\}_{n = 1}^{N - 1}$ \cite{Makri:ST, Makri:FP, Royer, Coalson}. Transforming between between these two sets of coordinates is linear \cite{FeynmanHibbs, Royer, Coalson}. Retaining the normalization constants $A$ from \eqref{freepartnorm}, the change in coordinates over integration is achieved with the determinant $J_{N - 1}$ of a Jacobian matrix \cite{Royer}. We then have
	\begin{equation}
		F(t_b - t_a) \equiv \lim_{N \to \infty} \frac{J_{N - 1}}{A} \int_{\mathbb R^{N - 1}} \! \left(\prod_{n = 1}^{N - 1} \frac{\diff a_n}{A}\right) \ex{i S[y] / \hbar}.
	\end{equation}
	Instead of first evaluating elements in the Jacobian matrix, Feynman and Hibbs leave it as an exercise to demonstrate for the free particle and harmonic oscillator Lagrangians under straight line segment time slicing that when $J_{N - 1}$ is collected with other terms from integration over each $a_n$, as $N \to \infty$ we find
	\begin{equation}
		J_{N - 1} \to (N - 1)! \left(\frac{\pi}{\sqrt{2}}\right)^{N - 1}\left(\frac{\epsilon}{t_b - t_a}\right)^{N / 2}.
		\label{sineJ}
	\end{equation}
	In the instance of the harmonic oscillator potential $V = m \omega^2 x^2 / 2$, after path integrating we then find the result
	\begin{equation}
    	F(t_b - t_a) = \left(\frac{m \omega}{i 2 \pi \hbar \sin \omega(t_b - t_a)}\right)^{1 / 2}.
    	\label{qhoF}
    \end{equation}
	
\section{Generalization of the time slicing method}

	We begin by partitioning the Lagrangian,
	\begin{equation}
		L = \Lo + L^{(1)} \Rightarrow S = S^{(0)} + S^{(1)},
		\label{genpart}
	\end{equation}
	where the Euler-Lagrange equation \eqref{E-L} for $\Lo$ results in an expression for the $\xcl$ satisfying the Dirichlet boundary conditions $\xcl(t_a) = x_a, \ \xcl(t_b) = x_b\ (S^{(0)}[b, a] = S^{(0)}[\xcl] = S_{cl}[b, a])$ and $L^{(1)}$ represents a term accounted for perturbatively by evaluating $L^{(1)}$ along $\xcl$. To assure minimization, a suitable choice in $\Lo$ is one where both $\Lo$ and $L$ satisfy the Legendre condition \eqref{Lgcond} over $\xcl$. Feynman and Hibbs consider the partition of kinetic energy $T$ and potential energy $V$,
	\begin{equation}
		\Lo = T, \qquad L^{(1)} = -V.
		\label{FHpart}
	\end{equation}
	For $T = m \dot x^2 / 2$, we then find with \eqref{E-L} over $\Lo$ the classical path
	\begin{equation}
		\xcl(t) = \frac{x_b (t - t_a) + x_a(t_b - t)}{t_b - t_a}.
		\label{E-Lsoln}
	\end{equation}
	Applying the time slicing partition \eqref{slicingconds} to this $\xcl$, we use the Heaviside step function $\Theta(t)$ to then express $x(t)$ and $\dot x(t)$ as
	\begin{align}
		\begin{split}
			x(t) = \; & \tfrac{x_1(t - t_a) + x_a(t_1 - t)}{\epsilon} \Theta(t_1 - t) + \sum_{n = 2}^{N - 1} \tfrac{x_n (t - t_{n - 1}) + x_{n - 1}(t_n - t)}{\epsilon} \Theta[(t - t_{n - 1})(t_n - t)] \\
			& + \tfrac{x_b (t - t_{N - 1}) + x_{N - 1}(t_b - t)}{\epsilon} \Theta(t - t_{N - 1}),
		\end{split}
		\label{freepathslice} \\
		\begin{split}
			\dot x(t) = \; & \tfrac{x_1 - x_a}{\epsilon} \Theta(t_1 - t) + \sum_{n = 2}^{N - 1} \tfrac{x_n - x_{n - 1}}{\epsilon} \Theta[(t - t_{n - 1})(t_n - t)] + \tfrac{x_b - x_{N - 1}}{\epsilon} \Theta(t - t_{N - 1}).
		\end{split}
	\end{align}
	The action is then again expressible as \eqref{actpart}. For each $S[n, n - 1]$, we again recover \eqref{shortactFH}, but now we find an explicit form for the average of the potential over the interval using substitution under integration,
	\begin{align}
		\bar V[n, n - 1] = & \; \frac{1}{\epsilon} \int_{t_{n - 1}}^{t_n} \! \diff t \, V\left(\tfrac{x_n (t - t_{n - 1}) + x_{n - 1}(t_n - t)}{\epsilon}\right)
		\label{avgV1} \\
		= & \; \frac{1}{x_n - x_{n - 1}} \int_{x_{n - 1}}^{x_n} \! \diff x \, V(x).
		\label{avgV2}
	\end{align}
	It can be demonstrated this integral average over the potential may be approximated by any of the potential averages $\bar V$ used by Feynman and Hibbs in \eqref{shortactFH} as $N\to\infty$. Because \eqref{avgV2} uses one less approximation than Feynman and Hibbs, Makri suggests a faster convergence to the propagator $K(x_b,t_b;x_a,t_a)$ when numerically evaluating path integrals should be achievable \cite{Makri:ST, Makri:FP}.
	
	The normalization constant $A$ in a short-time propagator \eqref{shortprop} found with the generalized time slicing method can be determined much in the same way as with \eqref{freepartnorm}. As a Green's function, a characteristic property of a propagator is that as time difference vanishes ($t_b - t_a \to 0^+$), the function approaches a Dirac delta function ($K(x_b, t_b; x_a, t_a) \to \delta(x_b - x_a)$). Another way to state this property is Green's functions are \textit{nascent delta functions}. Enforcing this property with \eqref{shortprop}, we find
	\begin{equation}
		\lim_{\epsilon \to 0^+} \frac{\ex{i S[n, n - 1] / \hbar}}{A} = \lim_{\epsilon \to 0^+}\frac{\ex{i S^{(0)}[n, n - 1] / \hbar}}{A} = \delta(x_n - x_{n - 1}).
	\end{equation}
	and therefore\footnote{Shankar states finding $A$ in this manner assumes it does not contain a dimensionless function $f(\epsilon)$ such that $f \to 1$ as $\epsilon \to 0^+$ \cite{Shankar}, yet there are instances like the damped harmonic oscillator described with the Lagrangian
	\[L(\dot x, x, t)=\frac{m}{2} \ex{2 \beta t}(\dot x^2 - \omega^2 x^2) \Rightarrow f(\epsilon) = \ex{\beta \epsilon}.\]
	The variable substitution $\xi(t) = x(t) \ex{\beta t}$ into this Lagrangian removes apparent violation of the assumption since $K(\xi_n, t_n; \xi_{n - 1}, t_{n - 1}) \to \delta(\xi_n - \xi_{n - 1})$, so $A$ can be determined despite dependence on the given $f$ \cite{UmYeonGeorge}.}
	\begin{equation}
		\lim_{\epsilon \to 0^+}\frac{1}{A} \int_{\mathbb R} \! \diff x_{n - 1} \, \ex{i S^{(0)}[n, n - 1] / \hbar} = 1.
		\label{S0norm}
	\end{equation}
	
	We are not restricted to the choice of $\Lo$ in \eqref{FHpart}. We could partition the potential as $V = V^{(0)} + V^{(1)}$ such that
	\begin{equation}
		\Lo = T - V^{(0)}, \qquad L^{(1)} = -V^{(1)}.
	\end{equation}
	For example, when we have a harmonic oscillator potential $V^{(0)} = m \omega^2 x^2 / 2$, for $\epsilon$ such that $\sin k \omega \epsilon \ne 0, \ k = 1, 2, \dotsc, N$ \cite{UmYeonGeorge, Horvathy}, we find in analogy to \eqref{freepathslice}
	\begin{align}
		\begin{split}
			x(t) = \; & \tfrac{x_1 \sin\omega(t - t_a) + x_a \sin\omega(t_1 - t)}{\sin\omega\epsilon} \Theta(t_1 - t) \\
			& + \sum_{n = 2}^{N - 1} \tfrac{x_n \sin\omega(t - t_{n - 1}) + x_{n - 1} \sin\omega(t_n - t)}{\sin\omega\epsilon} \Theta[(t - t_{n - 1})(t_n - t)] \\
			& + \tfrac{x_b \sin\omega(t - t_{N - 1}) + x_{N - 1} \sin\omega(t_b - t)}{\sin\omega\epsilon} \Theta(t - t_{N - 1}),
		\end{split}
	\end{align}
	with \eqref{S0norm} in analogy to \eqref{qhoF}
	\begin{equation}
		A = \frac{1}{F(\epsilon)} = \left(\frac{i 2 \pi \hbar \sin\omega\epsilon}{m \omega}\right)^{1 / 2},
	\end{equation}
	and if $V^{(1)} = V^{(1)}(x)$, we then find in analogy to (\ref{avgV1}, \ref{avgV2})
	\begin{align}
		\bar{V}^{(1)}[n, n - 1] = & \; \frac{1}{\epsilon} \int_{t_{n - 1}}^{t_n} \! \diff t \, V^{(1)} \left(\tfrac{x_n \sin\omega(t - t_{n - 1}) + x_{n - 1} \sin\omega(t_n - t)}{\sin\omega\epsilon}\right) \\
		= & \; \frac{\lvert \sin\omega\epsilon\rvert}{\omega\epsilon} \int_{x_{n - 1}}^{x_n} \! \diff x \, \frac{V^{(1)}(x)}{\sqrt{R^2 - x^2 \sin^2\omega\epsilon}},
	\end{align}
	where $R = R(x_n, x_{n - 1}) = \sqrt{x_n^2 + x_{n - 1}^2 - 2 x_n x_{n - 1} \cos\omega\epsilon}$.
	
	Ideally, an appropriate choice of $\Lo$ may result in a $\xcl$ which, under the time slicing procedure described in this section, produces a faster numerical convergence towards the propagator than would a straight-lined path.
	
\section{Generalization of the ``series" method}

	Notice equations (\ref{quadprop}, \ref{Fnorm}) imply the propagator can be represented as \cite{Shankar}
	\begin{equation}
		K(x_b, t_b; x_a, t_a) = \int_{y(t_a) = 0}^{y(t_b) = 0} \! \mathcal Dy \, \ex{i S[\xcl + y] / \hbar}.
		\label{seriesprop}
	\end{equation}
	We use this equation as a stepping stone to a generalization of the ``series" method, where we can find other series expressions to use for $y$ besides Fourier \eqref{Fourierseries}. Such functions will be characterized as satisfying the condition of minimizing the action $S$. Recall that if the action is to be minimized, it is necessary the value of the second variation \eqref{2ndvar} must be greater than zero and so the Legendre condition \eqref{Lgcond} must be satisfied. We will use these facts to derive such series.
	
	We again begin by partitioning the Lagrangian as in \eqref{genpart} such that if $S^{(0)}$ is minimized, then $S$ is minimized. We find the classical path $\xcl$ which minimizes $S^{(0)}$ such that both $\Lo$ and $L$ satisfy \eqref{Lgcond}. Let us now suppose $y$ can be expressed as a superposition of the set of eigenfunctions $\{u_n\}_{n=1}^\infty$,
	\begin{equation}
		y(t) = \sum_{n = 1}^\infty a_n u_n(t).
		\label{genseries}
	\end{equation}
	We determine these eigenfunctions with the second variation \eqref{preS-L2}, imposing the \textit{Sturm-Liouville equation} \cite{GelfandFomin, Schulman}
	\begin{equation}
		-\frac{\diff}{\diff t} \left(\frac{\partial^2 \Lo}{\partial \dotxcl^2} \dot u_n\right) + \left( \frac{\partial \ddot x_{cl}}{\partial \xcl} \frac{\partial^2 \Lo}{\partial \dotxcl^2} + \frac{\partial \dotxcl}{\partial \xcl} \frac{\partial^2 \Lo}{\partial \dotxcl \partial \xcl} \right) u_n = \lambda_n \frac{\partial^2 \Lo}{\partial \dotxcl^2} u_n.
		\label{St-Lv}
	\end{equation}
%	\begin{multline}
%		\left.\left[-\frac{\diff}{\diff t}\left(\dot u_n\frac{\partial^2}{\partial\dot x^2}\right)+u_n\left(-\frac{\diff}{\diff t}\frac{\partial^2}{\partial\dot x\partial x}+\frac{\partial^2}{\partial x^2}\right)\right]\Lo(\dot x,x,t)\right\rvert_{x=\xcl}\\
%		=\lambda_nu_n\Lo(\dotxcl,\xcl,t)
%	\end{multline}
	For all $x_{cl}$ determined from the Euler-Lagrange equation \eqref{E-L}, it can be demonstrated\footnote{For $L = L(\dot x, x, t)$, treating $(\dot x, x)$ as independent while differentiating with respect to $t$ \cite{GelfandFomin, Bernstein}, this follows from the identity
	\[\frac{\partial}{\partial x} \frac{\diff}{\diff t} \frac{\partial L}{\partial \dot x} = \frac{\diff}{\diff t} \frac{\partial^2 L}{\partial \dot x \partial x} + \frac{\partial \ddot x}{\partial x}\frac{\partial^2 L}{\partial \dot x^2} + \frac{\partial \dot x}{\partial x} \frac{\partial^2 L}{\partial \dot x \partial x}.\]}
	\begin{equation}
		-\frac{\diff}{\diff t} \frac{\partial^2 \Lo}{\partial \dotxcl \partial \xcl} + \frac{\partial^2 \Lo}{\partial \xcl^2}
		= \frac{\partial \ddot x_{cl}}{\partial \xcl} \frac{\partial^2 \Lo}{\partial \dotxcl^2} + \frac{\partial \dotxcl}{\partial \xcl} \frac{\partial^2 \Lo}{\partial \dotxcl \partial \xcl}.
	\end{equation}
	Solutions $\{u_n\}_{n=1}^\infty$ satisfy vanishing Dirichlet boundary conditions ($u_n(t_a) = u_n(t_b) = 0$) and the orthogonality condition
	\begin{equation}
		\int_{t_a}^{t_b} \! \diff t \, \frac{\partial^2 \Lo}{\partial \dotxcl^2} u_m u_n
		= \delta_{mn} \int_{t_a}^{t_b} \! \diff t \, \frac{\partial^2 \Lo}{\partial \dotxcl^2} u_n^2.
		\label{orthoncond}
	\end{equation}
%	\begin{equation}
%		\int_{t_a}^{t_b}\!\diff t\,u_m(t)u_n(t) \Lo(\dotxcl,\xcl,t)=S^{(0)}[\xcl]\delta_{mn}.
%	\end{equation}
%	We may then state something of an ``eigenfunctional" problem,
%	\begin{equation}
%		\delta^2S^{(0)}[\xcl;u_n]=\lambda_nS^{(0)}[\xcl],
%		\label{eigenfnl}
%	\end{equation}
%	permitting $\lambda_n<0$ when $S^{(0)}[\xcl]<0$.
	
	With equations (\ref{1stvar}, \ref{2ndvar}) being the first and second order variations of a functional $F=F[f]$, we can generalize to $n$th order with
	\begin{equation}
		\delta^n F[f_0; g] \equiv \left.\frac{\diff^n}{\diff \sigma^n} F[f_0 + \sigma g]\right\rvert_{\sigma = 0}.
		\label{nthvar}
	\end{equation}
	We may use this to write the action $S[\xcl + y]$ in \eqref{seriesprop} as
	\begin{align}
		S[\xcl + y] = & \sum_{n = 0}^\infty \frac{\delta^n S[\xcl; y]}{n!} \\
		= & \left.\exp\left(\frac{\diff}{\diff \sigma}\right) S[\xcl + \sigma y]\right\rvert _{\sigma = 0} \\
		= & \int_{t_a}^{t_b} \! \diff t \exp\left(\dot y \frac{\partial}{\partial \dotxcl} + y \frac{\partial}{\partial \xcl}\right) L(\dotxcl, \xcl, t).
		\label{acttrans}
	\end{align}
	Using for $y$ our expression \eqref{genseries} and $\{u_n\}_{n = 1}^\infty$ satisfying (\ref{St-Lv}, \ref{orthoncond}), we may find with (\ref{acttrans}, \ref{actexpand}) that
	\begin{equation}
		S^{(0)}[\xcl + y] = \int_{t_a}^{t_b} \! \diff t \, \Lo \left(\dotxcl + \sum_{n = 1}^\infty \sqrt{\lambda_n} a_n u_n, \xcl, t\right),
		\label{eigenshift}
	\end{equation}
%	\begin{equation}
%		S^{(0)}[\xcl+y]=\int_{t_a}^{t_b}\!\diff t\exp\left(\sum_{n=1}^\infty\sqrt{\lambda_n} a_n u_n(t)\right)\Lo(\dotxcl,\xcl,t),
%	\end{equation}
	provided the subsidiary condition
	\begin{equation}
		\int_{t_a}^{t_b} \! \diff t \, \frac{\partial \Lo}{\partial \dotxcl} u_n = 0
		\label{addcon}
	\end{equation}
%	\begin{equation}
%		\int_{t_a}^{t_b}\!\diff t\,u_n(t)\Lo(\dotxcl,\xcl,t)=0
%	\end{equation}
	can be satisfied so that $\delta S^{(0)}[\xcl; y] = 0$. This can restrict the possible boundary condition values taken at $\xcl(t_a)$ and $\xcl(t_b)$.%This condition constrains the eigenvalue spectrum of $y$ effectively to that of a Lagrangian $L = L(\dot x, t)$. It occurs under resonance when the eigenvalue spectrum consists of eigenvalues $\lambda_i, \, \lambda_j$ satisfying the relation $\lambda_j = \lambda_i - \partial \ddot x_{cl} / \partial \xcl$. %\footnote{When $\Lo$ is a $k$th-order polynomial in $(\dot x,x)$, attempting to express $S^{(0)}$ with \eqref{eigenshift} imposes additional constraints similar to \eqref{addcon} for $l$th-ordered products of $\{u_n\}_{n=1}^\infty$ since $\delta^lS^{(0)}[\xcl;y]=0\ \forall\ l>k$ using \eqref{nthvar}. Attempting to satisfy these additional constraints as $l\to\infty$ may remove all nonzero eigenvalues from the spectrum of $y$. Directly substituting \eqref{genseries} into either (\ref{actexpand}, \ref{acttrans}) and Taylor expanding, we should not have to constrain the eigenvalue spectrum with \eqref{addcon} or any of these other additional constraints. But we lose the exponential factorablity of \eqref{eigenshift} when we apply this to non-polynomial $\Lo$.}
	
	As an example of the generalized series method, consider when $\Lo = \Lo(\dot x)$ such that $\dotxcl$ is constant. With (\ref{St-Lv}, \ref{orthoncond}, \ref{addcon}), we find eigenfunctions
	\begin{equation}
		u_n(t) = \sin\left(2 n \pi \frac{t - t_a}{t_b - t_a}\right) = -\sin\left(2 n \pi \frac{t_b - t}{t_b - t_a}\right).
		\label{freeparteigenfns}
	\end{equation}
%	\begin{equation}
%		u_n(t)=\sqrt2\sin\left(2\pi n\frac{t-t_a}{t_b-t_a}\right)=-\sqrt2\sin\left(2\pi n\frac{t_b-t}{t_b-t_a}\right).
%	\end{equation}
	%Another example but contrasting in that $\Lo=\Lo(\dot x,x)$ and in being a polynomial, when $\Lo$ is that of a harmonic oscillator we find \textit{Mathieu sine functions} with phase-shifted argument for $\{u_n\}_{n=1}^\infty$ \cite{JahnkeEmde}.
	As another example, when $\Lo$ is that of a harmonic oscillator with angular frequency $\omega$, it may at first appear we would have the exact same set of eigenfunctions as \eqref{freeparteigenfns}. However, \eqref{addcon} cannot be satisfied unless we have with $\xcl$ that $x_b = x_a\ (-x_a)$ for odd (even) $n$. Furthermore, the corresponding eigenvalues $\{\lambda_n\}_{n = 1}^\infty$ initially determined with \eqref{St-Lv} are shifted by $\omega^2$,
	\begin{equation}
		\lambda_n = \left(\frac{n \pi}{t_b - t_a}\right)^2-\omega^2.
	\end{equation}
	Since we must have $\lambda_n>0\ \forall \ n = 1, 2, \dotsc$ in order for $S^{(0)}[\xcl]$ to correspond to a minimum, we are constrained to $n > \omega (t_b - t_a) / \pi$, inclusive of both odd and even $n$ if $x_b \ne \pm x_a$.
	
	Comparing to the Jacobian determinant $J_{N - 1}$ \eqref{sineJ} from the Fourier sine series \eqref{Fourierseries} used in linearly transforming $\{y_n\}_{n = 1}^{N - 1} \mapsto \{a_n\}_{n = 1}^{N - 1}$ as $N \to \infty$, we find with \eqref{genseries} and for comparison's sake with \eqref{freeparteigenfns} that
	\begin{equation}
		J_{N-1}\to(N-1)!\left(\frac{2\pi}{\sqrt2}\right)^{N-1}\left(\frac{\epsilon}{t_b-t_a}\right)^{N/2}.
		\label{sine2J}
	\end{equation}
%	\begin{equation}
%		J_{N-1}\to(N-1)!(2\pi)^{N-1}\left(\frac{\epsilon}{t_b-t_a}\right)^{N/2}.
%	\end{equation}
	Derivation of this result is straightforward due to separability between the action over $\xcl$ and $y$ in \eqref{quadprop} \cite{FeynmanHibbs}. If \eqref{seriesprop} cannot be separated as such, explicit evaluation of $J_{N - 1}$ may be necessary. We present here two methods similar to those of Royer and Coalson \cite{Makri:ST, Makri:FP, Royer, Coalson}, both of which start with the assumption that we approximate \eqref{genseries} to the first $N-1$ terms,
	\begin{equation}
		y(t) \simeq \sum_{n = 1}^{N - 1} a_n u_n(t).
		\label{genseriesapprox}
	\end{equation}
	
\begin{figure}[H]
		\centering
		\includegraphics[scale = 0.7]{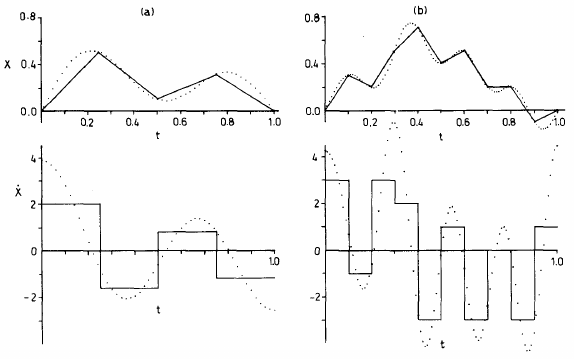}
		\caption{Transforming from $\{x_n\}_{n = 1}^{N - 1} \ (x_0 = x_N = 0)$ to $\{a_n\}_{n = 1}^{N - 1}$ over $t\in[0, 1]$. Fitting points from a time sliced path of $N$ straight line segments (solid line) using a $N$-term Fourier sine series (dotted line), for (a) $N = 4$ and (b) $N = 10$. The top row represents the path $x$ and the bottom row its time derivative $\dot x$ \cite{Royer}.}
		\label{slicetoseriesfit}
	\end{figure}

	The first method is to approximately evaluate $\{y_n\}_{n = 1}^{N - 1}$ in terms of \eqref{genseriesapprox} using the time slicing condition \eqref{slicingconds},
	\begin{equation}
		y_n = y(t_n) \simeq \sum_{m = 1}^{N - 1} a_m u_m(t_n),\qquad n = 1, 2, \dotsc, N - 1.
		\label{slicetoseries}
	\end{equation}
	Taking the determinant of the matrix transforming $\{y_n\}_{n = 1}^{N - 1} \mapsto \{a_n\}_{n = 1}^{N - 1}$ is then $J_{N - 1}$. For example, still using \eqref{freeparteigenfns} for comparison to \eqref{Fourierseries}, we have
	\begin{equation}
		y_n \simeq \sum_{m = 1}^{N - 1} a_m \sin\left(\frac{2 n m \pi}{N}\right).
		\label{freepartslicetoseries}
	\end{equation}
%	\begin{equation}
%		y_n\simeq\sqrt2\sum_{m=1}^{N-1}a_m\sin\left(\frac{2n m\pi}{N}\right).
%	\end{equation}
	Convergence to \eqref{sine2J} from \eqref{freepartslicetoseries} might be slow, as the factors which emerge in \eqref{sine2J} are not readily apparent unless the approximation $\sin x\simeq x$ for small $x$ as $N \to \infty$ can be somehow applied to \eqref{freepartslicetoseries} in a consistent manner. Furthermore, as demonstrated in the bottom row of figure \ref{slicetoseriesfit}, $\dot y$ found with this fitting can take on values over $t\in[t_a, t_b]$ considerably different between the time sliced and series paths. This may slow the numerical convergence towards the propagator when compared to the time slicing method \cite{Makri:FP, Royer, Coalson}.
	
	The second method is to make use of the orthogonality of $\{u_n\}_{n = 1}^{N - 1}$ from \eqref{orthoncond}. We have
	\begin{equation}
		\int_{t_a}^{t_b} \! \diff t \, \frac{\partial^2 \Lo}{\partial \dotxcl^2} u_n y = a_n \int_{t_a}^{t_b} \! \diff t \, \frac{\partial^2 \Lo}{\partial \dotxcl^2} u_n^2,
		\label{seriestoslice}
	\end{equation}
%	\begin{equation}
%		a_n=\frac{1}{S^{(0)}[\xcl]}\int_{t_a}^{t_b}\!\diff t\,y(t)u_n(t) \Lo(\dotxcl,\xcl,t),
%	\end{equation}
	so expressing $y(t)$ in terms of $\{y_n\}_{n = 1}^{N - 1}$ we can construct the matrix transforming $\{a_n\}_{n = 1}^{N - 1} \mapsto \{y_n\}_{n = 1}^{N - 1}$. The determinant of this matrix is then  $1 / J_{N - 1}$. Continuing with the example of \eqref{freeparteigenfns}, using \eqref{seriestoslice} with the corresponding time sliced partition \eqref{freepathslice} we find
	\begin{equation}
		a_n = \frac{2 N}{n^2 \pi^2} \sin^2 \bigg(\frac{n \pi}{N}\bigg) \sum_{m = 1}^{N - 1} y_m \sin\left(\frac{2 n m \pi}{N}\right).
		\label{freepartseriestoslice}
	\end{equation}
	Given the difference in prefactor between \eqref{freepartslicetoseries} and \eqref{freepartseriestoslice}, using the second method may result in a $J_{N - 1}$ which converges to \eqref{sine2J} quicker than when using the first method. This second method may therefore improve the convergence rate towards the propagator when numerically evaluating the path integral with the series method.\footnote{A possible critique relating to constraint \eqref{addcon} could be that when transforming between $\{y_n\}_{n = 1}^{N - 1} \leftrightarrow \{a_n\}_{n = 1}^{N - 1}$ using either (\ref{slicetoseries}, \ref{seriestoslice}), not all $y_n$ can be expressed in terms of $a_n$ for some finite $N$, and vice versa. For example, when $N = 2 k$ such that $k = 1, 2, \dotsc$, with \eqref{freepartslicetoseries} we find $y_k = 0$ and with \eqref{freepartseriestoslice} we find $a_k = 0$, among other problems. In this instance we should take this to imply only consideration of $N = 2 k + 1$. We must remind ourselves (\ref{freepartslicetoseries}, \ref{freepartseriestoslice}) are only properly applicable as $N \to \infty$, so we can have $2 k + 1 = N \to \infty$ if we instead first assume $k \to \infty$. This just demonstrates we have to be mindful of our limiting procedures. Furthermore, we can always choose generally which $u_n$ are used to approximately construct $y$, not just those in \eqref{genseriesapprox}. Selecting which $a_n$ to represent $\{y_n\}_{n = 1}^{N - 1}$ may address similar problems with other series besides (\ref{freepartslicetoseries}, \ref{freepartseriestoslice}).}
	
\section{Discussion}

    Portions of both generalized methods have been demonstrated in literature \cite{Shankar, GelfandFomin, Schulman, Makri:ST, Makri:FP, UmYeonGeorge, Horvathy, Davison, deCarvalhoCavalcanti}. Early on Davison deduced series other than Fourier could be used \cite{Davison}. Investigation into the series method was inspired by the work of Gelfand and Fomin where it is assumed $\int_{t_a}^{t_b} \! \diff t \, u_mu_n = \delta_{mn}$ such that $\delta^2 S^{(0)}[\xcl, u_n] = \lambda_n > 0$ can be inferred as opposed to \eqref{orthoncond} \cite{GelfandFomin, Schulman, Horvathy}. A method sharing similarities to the generalized series method presented here is in using the method of steepest descent, or stationary phase approximation, to provide an altenative method in approaching the Jacobian determinant found under (\ref{sineJ}, \ref{sine2J}) using a determinant property of multivariate Gaussians \cite{Schulman, Ranfagni, ItzyksonZuber}.

    Hopefully, what has been written here will be of some utility. If not with propagators in quantum mechanics, then perhaps with ghost fields coming from Lagrangian densities $\mathcal L = \mathcal L(\partial_\mu \phi, \phi, x^\mu)$ in quantum field theory \cite{ItzyksonZuber, Ryder}. What may be an interesting question to then ask is if the resulting field operators coming from $\mathcal L$ and analogous to $\{u_n\}_{n = 1}^\infty$ have physical meaning.
    
    %field operators analogous to $\{u_n\}_{n = 1}^\infty$, solved from a system of Sturm-Liouville equations \eqref{St-Lv} derived using the field operators $(\partial_\mu \phi, \phi)$ solved from $\mathcal L$ and with vanishing Dirichlet boundary conditions on $x^\mu$.
	
\section*{Acknowledgements}

	The author would like to thank Dr. John Learned for providing the motivation to write the paper's initial draft, Steven Smith for his assistance in revising the paper, and Dr. Xerxes Tata for both his revisions and discussions on the paper's materials and ideas.
	
\bibliographystyle{unsrt}
\bibliography{path-integral}

\end{document}